\documentclass[sigconf]{acmart}

\usepackage{booktabs} 
\usepackage{multirow}
\usepackage[colorinlistoftodos]{todonotes}

\author{Katie Lim}
\authornote{This work done at Princeton University}
\affiliation{%
    \institution{University of Washington}
}
\email{katielim@cs.washington.edu}

\author{Jonathan Balkind}
\affiliation{%
    \institution{Princeton University}
}
\email{jbalkind@cs.princeton.edu}

\author{David Wentzlaff}
\affiliation{%
    \institution{Princeton University}
}
\email{wentzlaf@princeton.edu}

\begin{document}
\title{JuxtaPiton: Enabling Heterogeneous-ISA Research with RISC-V and SPARC FPGA Soft-cores}

\begin{abstract}
Energy efficiency has become an increasingly important concern in computer architecture due to the end of Dennard scaling. Heterogeneity has been explored as a way to achieve better energy efficiency and heterogeneous microarchitecture chips have become common in the mobile setting. 

Recent research has explored using heterogeneous-ISA, heterogeneous microarchitecture, general-purpose cores to achieve further energy efficiency gains. However, there is no open-source hardware implementation of a heterogeneous-ISA processor available for research, and effective research on heterogeneous-ISA processors necessitates the emulation speed provided by FPGA prototyping.
This work describes our experiences creating JuxtaPiton by integrating a small RISC-V core into the OpenPiton framework, which uses a modified OpenSPARC T1 core. This is the first time a new core has been integrated with the OpenPiton framework, and JuxtaPiton is the first open-source, general-purpose, heterogeneous-ISA processor. JuxtaPiton inherits all the capabilities of OpenPiton, including vital FPGA emulation infrastructure which can boot full-stack Debian Linux. Using this infrastructure, we investigate area and timing effects of using the new RISC-V core on FPGA and the performance of the new core running microbenchmarks.
\end{abstract}

%
%


\maketitle

\section{Introduction}
Energy efficiency has become an increasingly important concern for modern processors. The end of Dennard scaling means that power dissipation no longer decreases with feature length. This has resulted in new research problems such as dark silicon \cite{dark_silicon_taylor} that require a new emphasis on energy efficiency.
Additionally, demands in both the datacenter and the mobile setting have made power and energy efficiency more important. Datacenters now account for several percent of global energy usage \cite{datacenter_energy_usage}. As we increasingly rely on datacenters for computation, energy efficiency becomes more important both for economic and environmental reasons \cite{energy_prop_compute}. In a mobile setting with limited cooling and demand for better battery life, energy efficiency cannot be avoided.

\begin{figure}
\centering
\includegraphics[width=\linewidth]{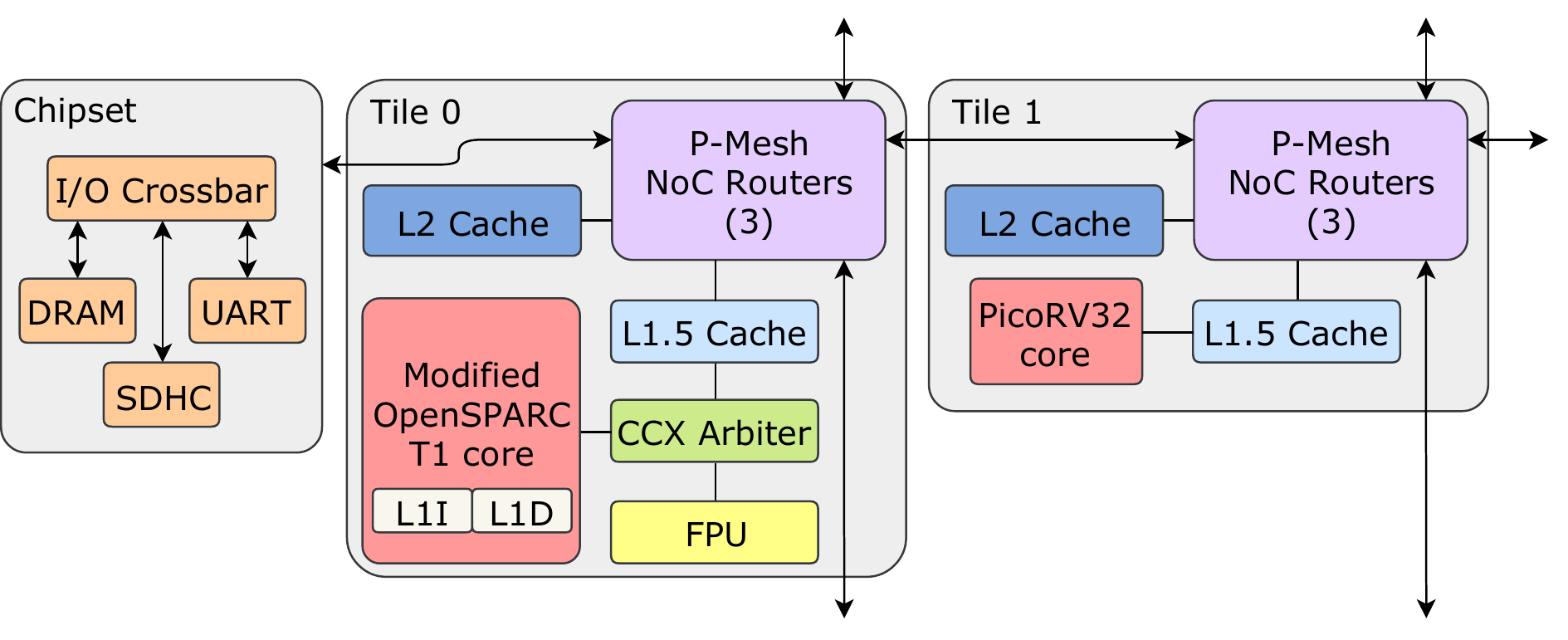}
\vspace{-.4cm}
\caption{Architecture of an OpenSPARC T1 tile and a PicoRV32 tile connected to memory. (Derived from \cite{openpiton_asplos})}
\label{fig:pico_mem_hierarchy}
\vspace{-.6cm}
\end{figure}

In response to the need for energy efficiency, research has explored introducing heterogeneity into processors. Heterogeneous processors seek to cater to the fact that not all applications have the same computational demands.

Processors with heterogeneous microarchitecture have become common in modern cellphones for their energy efficiency benefits. Examples include ARM's big.LITTLE architectures \cite{arm-big.LITTLE} and Apple's A11 processor \cite{apple-a11}. These processors use smaller, lower power cores for applications that do not need the computational power of larger, more power hungry cores to achieve better energy efficiency without significantly impacting the user experience.

Recent research by Venkat et al. explored using general-purpose cores with heterogeneous microarchitecture and heterogeneous ISA in simulation \cite{venkat-hetero-chip}. They found potential energy and performance benefits from this type of architecture. However, further work on this topic has been difficult due to a lack of suitable prototyping platforms which can explore the implications of heterogeneous-ISA processor designs in a full-stack system at speeds suitable for rapid prototyping

To assist heterogeneous-ISA research, we have created JuxtaPiton by integrating PicoRV32 \cite{pico}, a small RISC-V core, into the OpenPiton framework \cite{openpiton_asplos}, an open-source manycore research platform which uses a modified OpenSPARC T1 core \cite{opensparc_t1}. This is the first time another core has been integrated into OpenPiton, and we believe the JuxtaPiton FPGA platorm is \textbf{the first open-source implementation of a general-purpose, heterogeneous-ISA processor}. Given that heterogeneous ISA architecture is an emerging area of research that requires investigation of issues in hardware and software, an FPGA implementation will prove the most helpful to heterogeneous-ISA researchers. Architects will be able to modify any aspect of the design and prototype their research ideas on FPGA. At the same time, OS researchers can run complex, full-stack software on an FPGA-speed hardware system to evaluate their designs. 

Heterogeneous-ISA processors pose interesting challenges to software. Some prior research has been done in simulation \cite{DeVuyst-exe-migration}. However, OS research projects such as Popcorn Linux \cite{popcorn_os_multi_isa} or K2 \cite{k2_os} have relied on hardware platforms that do not support shared memory. While this is a common design point, shared memory systems are easier to program and common in homogeneous-ISA processors, like those used in mobile systems on chip (SoCs). Additionally, hardware shared memory is needed for efficient process migration between ISAs \cite{DeVuyst-exe-migration,popcorn_mars}. However, it is difficult to build a heterogeneous-ISA, shared memory processor using off-the-shelf parts. In JuxtaPiton, the PicoRV32 core has fully cache coherent shared memory with the SPARC core enabled by OpenPiton's P-Mesh cache system. 

This paper details our experience using OpenPiton and RISC-V to create JuxtaPiton. By leveraging the OpenPiton infrastructure, we were able to quickly construct a functional system and implement software infrastructure to run static C binaries on the PicoRV32 core hosted by the OpenSPARC T1 core, and the PicoRV32 core is able to proxy syscalls to the OpenSPARC T1 core. We also evaluate some of the trade-offs of the PicoRV32 core compared to the OpenSPARC T1 core. We look at potential area and timing improvements as well as the performance impact of using the simpler PicoRV32 core instead of the OpenSPARC T1 core. 

\section{Architecture}
\begin{table}[t]
\begin{tabular}{|c|c|c|}
\hline
\cellcolor{gray}  	& \textbf{OpenSPARC T1}	& \textbf{PicoRV32}	\\
\hline
\textbf{ISA}       	& SPARC v9				& RISC-V I		  	\\
\hline
\textbf{Word size} 	& 64 bits		  		& 32 bits			\\
\hline
\textbf{Endianness} & Big endian			& Little endian		\\
\hline
\multirow{2}{*}{\textbf{Implementation}} & 6-stage in-order & \multirow{2}{*}{{Multicycle}} \\
& pipeline & \\
\hline
\textbf{L1 Cache}  	& Yes					& No				\\
\hline
\textbf{MMU/TLB}		& Yes					& No				\\
\hline
\textbf{FPU}		& Yes					& No				\\
\hline
\textbf{Privileged Mode} & Yes				& No				\\
\hline
\end{tabular}
\caption{A summary of the major differences between the OpenSPARC T1 core and the PicoRV32 core}
\label{tab:core_comp}
\vspace{-.7cm}
\end{table}

To construct our framework, we leverage the PicoRV32 and the OpenPiton framework. We integrated these two open-source projects by connecting PicoRV32 to the OpenPiton cache hierarchy. These cores were chosen because they are very different. Table \ref{tab:core_comp} summarizes differences between the two cores.

\subsection{OpenPiton}
OpenPiton has a tiled manycore architecture. Each tile has a core, an FPU, three P-Mesh NoC routers, and caches. The original OpenPiton core is a modified OpenSPARC T1 core implementing the SPARCv9 ISA, and has a 6-stage in-order pipeline. The core has an instruction cache and a data cache. Each tile also has two levels of cache: the L1.5 and the L2. The L1.5 cache is equivalent to a private L2 cache, and the L2 cache is equivalent to a shared, distributed Last Level Cache (LLC). An OpenSPARC tile is shown in Figure \ref{fig:pico_mem_hierarchy} as Tile 0. Cache coherence is maintained using OpenPiton's cache coherence protocol, P-Mesh. The OpenPiton framework supports running designs in simulation as well as implementing designs for FPGA.

\subsection{PicoRV32}
PicoRV32 is a multicycle implementation of RV32I, the 32-bit core RISC-V ISA. It has no caches, and it does not support virtual memory. The core also does not implement the RISC-V privileged specification, so it is hosted by the OpenSPARC T1 core. 

We decided on the PicoRV32 core for several reasons. First, the PicoRV32 core is open-source and is written in synthesizable Verilog RTL. It also has been applied in a number of settings by the community and has been the subject of formal verification \cite{pico_formal}. Second, its simpler microarchitecture meant we would have a heterogeneous-ISA, heterogeneous microarchitecture system. We chose to use a core with a vastly simpler microarchitecture compared to the OpenSPARC T1 in order to research differences in microarchitecture.
\vspace{-.2cm}

\subsection{Integration}\label{sec:integration}
To integrate the PicoRV32 core into the OpenPiton infrastructure, we connected it behind OpenPiton's L1.5 cache by adding transducers that convert the memory requests from the PicoRV32 core to OpenPiton L1.5 cache operations. 
This creates a tile where the OpenSPARC T1 core is replaced by the PicoRV32 core. We also removed the FPU from the tile with the PicoRV32 core, since the core does not support the RISC-V floating point extension. A diagram of an OpenSPARC T1 tile and a PicoRV32 tile connected to each other and memory is shown in Figure \ref{fig:pico_mem_hierarchy}. 

Connecting the PicoRV32 core to the L1.5 cache enables it to use the P-Mesh cache coherence protocol without modification to any existing infrastructure, therefore OpenSPARC T1 cores and PicoRV32 cores can share memory. Because interrupts traverse the caches, connecting the PicoRV32 core to the L1.5 cache also enables the PicoRV32 core to receive interprocessor interrupts from the OpenSPARC T1 core. The PicoRV32 core is brought out of reset using an interrupt sent from the OpenSPARC T1 core.

When we integrated the PicoRV32 core, we had to consider byte endianness differences, because SPARC is a big endian ISA whereas RISC-V is a little endian ISA. We chose to flip the outgoing and incoming data buses for the PicoRV32 core, such that the data accessed by the PicoRV32 core is stored little endian in memory. 
SPARCv9 does support little endian data accesses with use of special assembly instructions, but in our higher-level C code, we choose to use endian-flipping macros when interacting with data for the PicoRV32 cores.

The PicoRV32 core interacts with the L1.5 cache slightly differently than the OpenSPARC T1 core does. This is because the PicoRV32 core does not have an L1 cache, so the L1.5 is the PicoRV32 core's first-level cache whereas the OpenSPARC T1 core has L1 caches, making the L1.5 cache the OpenSPARC T1 core's second-level cache.

To improve performance, we chose to have the L1.5 cache instructions and data for the PicoRV32 core. This is in contrast with the OpenSPARC T1 core where instructions are not cached in the L1.5 cache, but only in the L1 and the L2 caches. 

Second, OpenPiton's L1 cache is write-through, so writes must always go to the L1.5 cache when using the OpenSPARC T1 core. However, the L1 cache is the same associativity and capacity as the L1.5 cache, so any read that would hit in the L1.5 cache also hits in the L1 cache. This means that reads do not go to the L1.5 cache. However, when using the PicoRV32 core, both reads and writes will go to the L1.5 cache.

We found that modifying the OpenPiton framework to fit our needs was quick and the changes were relatively minimal even though we were replacing a core which is a relatively major change. The changes were isolated to the transducers and slight modifications to core and tile instantiations as well as instantiations within the tile. so we could select between OpenSPARC T1 and PicoRV32 cores. Once the core was integrated, we were able to instantiate multiple tiles with different cores without further modification to the infrastructure. 

JuxtaPiton maintains all the functionality of the original OpenPiton framework. For example, the PicoRV32 core is able to access all of the I/O devices from the original OpenPiton framework including the SD card and the UART. We were also able to have the PicoRV32 core read instructions straight from the SD card and write characters to the UART during testing and evaluation.

Furthermore, because the OpenPiton framework contains push-button scripts for implementation on FPGA, and the PicoRV32 core was synthesizable, we could immediately test our design on FPGA, and we used the FPGA to efficiently prototype JuxtaPiton. Being able to implement JuxtaPiton on FPGA was also crucial to our software prototyping, which we describe in more detail in Section \ref{sec:software}. 

\section{Software Support}\label{sec:software}
We are able to run the RISC-V assembly test suite and statically-linked C binaries on the PicoRV32 core. We augmented the OpenPiton simulation infrastructure to enable us to compile and run RISC-V assembly tests by adding a script to generate the proper memory image for the PicoRV32 core. The PicoRV32 core integrated in the OpenPiton framework is able to run and pass all of the RV32UI tests from the official \texttt{riscv-tests} distribution \cite{riscv_tests}. 

For evaluation, we also built software infrastructure to enable the PicoRV32 core to run statically-linked C binaries hosted by the OpenSPARC T1 core. The OpenSPARC T1 core is responsible for loading the binaries into memory and proxying any syscalls from the PicoRV32 core. 

The OpenPiton software stack consists of full-stack Debian Linux and a lightly modified version of the OpenSPARC T1 hypervisor. To support the PicoRV32 core, we wrote a userspace proxy program to load binaries and proxy syscalls for the PicoRV32 core to OpenSPARC T1 core running full-stack Debia Linux. This enables the PicoRV32 core to access OS resources, such as the file system, transparently. We also added two new Linux syscalls, and added a new hypercall.

\begin{figure}[t]
\includegraphics[width=3.25in]{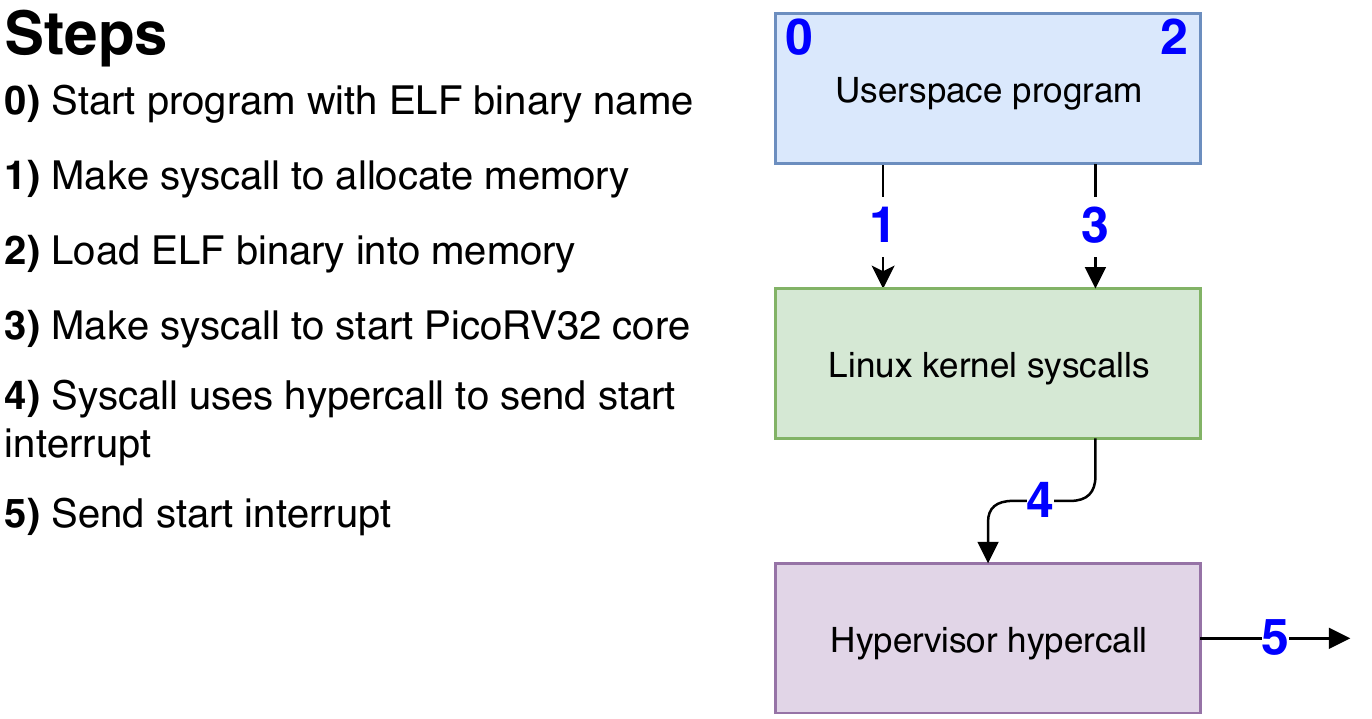}
\caption{Flowchart of the process used to load and start running a binary for the PicoRV32 core}
\label{fig:load_binary}
\vspace{-.6cm}
\end{figure}

An overview of the process to run a binary on the PicoRV32 core is shown in Figure \ref{fig:load_binary}. The userspace proxy program on the OpenSPARC T1 core with the name of the binary to be run provided as a command line argument (Step 0). The added syscall \texttt{pico\_setup} is then used to allocate a region of physical memory for the PicoRV32 core (Step 1). Once the syscall returns, the binary is loaded into the allocated memory (Step 2). After the binary is completely loaded, the OpenSPARC T1 core send the start interrupt to the PicoRV32 core using the new syscall \texttt{pico\_start} (Step 3). The syscall then calls the new hypercall \texttt{hycall\_pico\_start} (Step 4). Finally, the hypervisor sends the start interrupt to the PicoRV32 core (Step 5). 

The userspace proxy program continues running to proxy syscalls from the PicoRV32 core. We take advantage of the fact that the OpenSPARC T1 core is running full-stack Linux and have the OpenSPARC T1 core perform syscalls. Binaries that run on the PicoRV32 core are linked against a version of Newlib where the syscall stubs are modified to write the syscall number and arguments out to the shared piece of memory. The OpenPiton userspace program polls on this memory and when it sees the PicoRV32 core needs a syscall serviced, it reads the number and arguments out of memory and makes the syscall in Linux itself. To return the result, the OpenSPARC T1 core writes it back to the shared memory region. The PicoRV32 core can then read this result and make use of it.

We host the PicoRV32 core since it does not implement the RISC-V privileged specification, but a core with a privileged specification implementation could also be hosted using the same setup. Additionally, system software support for heterogeneous-ISA systems with self-hosting cores in a shared memory system is an active area of research.

When developing our software infrastructure, having JuxtaPiton running on FPGA was crucial. When running on FPGA, the SPARC core is able to run Linux, which is practically impossible in behavioral simulation due to the orders of magnitude difference in speed of simulation (tens of kilohertz) versus FPGA emulation speed (ten of megahertz). Linux provides a much more fully featured enviroment for developing software, which enabled us to develop a RISC-V ELF binary loader on the SPARC core. Additionally, Linux is required for PicoRV32 to be able to proxy syscalls.

\section{Evaluation}
All evaluation was done using a Digilent Genesys2 FPGA board using Xilinx Vivado 2015.4 to implement designs for the boards. 

\subsection{Area Analysis}
For our area analyses we used Xilinx Vivado to build bitfiles for the Genesys2. The Genesys2 uses the Xilinx Kintex-7 FPGA (XC7K325T-2FFG900C) \cite{g2_specs}. We used Vivado's default synthesis strategy and the \texttt{phys\_opt\_design} implementation strategy. 

We looked at building designs with 1 OpenSPARC T1 tile or 1 PicoRV32 tile at frequencies between 50 MHz and 100 Mhz. We saw no significant change in resource utilization for either an OpenSPARC T1 tile or a PicoRV32 tile when increasing frequency. 

There is a slight gain in maximum frequency when using the PicoRV32 core over the OpenSPARC T1 core. The maximum frequency that meets timing is 109.091 Mhz for an OpenSPARC T1 tile and 114.286 MHz for a PicoRV32 tile. For the OpenSPARC T1 tile, the critical path is in the D-TLB. For the PicoRV32 tile, the critical path is in the L2 cache.

\begin{table}[h]
\begin{tabular}{|c|c|c|c|}

\hline
\textbf{Tile Type}	& \textbf{Core LUTs} & \textbf{Tile LUTs} & \textbf{Core BRAMs}\\
\hline
\hline
\textbf{OpenSPARC T1}	& 36756				 & 64695	    	  & 24				   \\
\hline
\textbf{PicoRV32}	& 1076				 & 21862 			  & 0				   \\
\hline

\end{tabular}
\caption{Resource utilization for OpenSPARC T1 and PicoRV32 cores and tiles on the Xilinx Kintex-7 FPGA}
\label{tab:utilization}
\vspace{-.6cm}
\end{table}

We did see a significant area improvement gained by using the PicoRV32 core over the OpenSPARC T1 core. Utilization is shown in Table \ref{tab:utilization}. We found that a PicoRV32 core uses approximately $\frac{1}{30}th$ the look-up tables (LUTs) of an OpenSPARC T1 core. The resulting PicoRV32 tile uses approximately $\frac{1}{3}rd$ the LUTs of an OpenSPARC T1 core. We can fit 2 OpenSPARC T1 cores on the Genesys2 or 7 PicoRV32 cores. In both cases, the limiting resource is the LUTs.

\subsection{Memory Hierarchy Latency}
We also investigated the latency in cycles to different parts of the memory hierarchy to better understand performance of the PicoRV32 core. Cycle counts for other instructions that are only dependent on the core itself are already available on the GitHub page for the core.

We measured latency of using a memory operation between two \texttt{rdcycle} instructions. The results are shown in Table \ref{table:mem_latency}. The raw measurements given are the difference between the cycle counts before and after a memory operation. 
However, the PicoRV32 core is unpipelined and takes multiple cycles to execute each instruction. As such, the raw measurements must be adjusted to gain more insight into how much of the latency is actually from operations within the memory hierarchy versus the time for the other portions of the instruction to execute.

We first looked at determining the true L1.5 cache hit time from the raw L1.5 cache hit time of 17 cycles. Every instruction requires at least one cache access to the L1.5 cache to fetch it, so a memory instruction is actually two accesses to the L1.5 cache. Some of the latency is also from fetching the second \texttt{rdcycle} instruction. Thus, the instructions used to measure memory latency account for 3 accesses to the L1.5 cache.
In addition, load and store instructions take 5 cycles in the PicoRV32 core. This was given by the official documentation for the PicoRV32 core and verified in simulation.
For accesses to the L1.5 cache, after subtracting off the 5 cycles for the memory instruction and dividing by the 3 memory hierarchy accesses, we get that an L1.5 cache hit for the PicoRV32 takes \textbf{4 cycles}.
The time for a DRAM is much higher at around 113 cycles, and there is some slight variance in the measurements. This is most likely due to the fact that requests that go all the way to DRAM cross clock domains and need to go through asynchronous FIFOs. Depending on when the request gets to the FIFO relative to the other clock domain, there may be variance in the number of cycles it is waiting.
In this part of the test, both instruction reads go to the L1.5 cache and only the actual memory access goes all the way out to memory. With this in mind, one operation to memory for the PicoRV32 core is about \textbf{100 cycles}. Latency from L1.5 cache to L2 cache is the same for PicoRV32 and OpenSPARC cores and varies with core count and the L2 cache homing policy as described in \cite{openpiton_asplos}.


\begin{table}[t]
\begin{tabular}{|c|c|c|c|c|}
\hline
\multirow{2}{*}{\textbf{Operation}} & \multicolumn{2}{|c|}{\textbf{Measured}} & \multicolumn{2}{|c|}{\textbf{True}} \\
& \multicolumn{2}{|c|}{\textbf{Latency (cycles)}} & \multicolumn{2}{|c|}{\textbf{Latency (cycles)}} \\

\hline
\hline
\cellcolor{gray} & \textbf{Cached} & \textbf{Uncached}  & \textbf{Cached} & \textbf{Uncached} \\
\hline
\textbf{Load}	 & 17   		   &	$113	\pm	 1$ & 4   		   	  &	$100	\pm	 1$   \\
\hline	  
\textbf{Store}	 & 17	  		   &	$113	\pm 1$  & 4   		      & $113	\pm  1$   \\
\hline
\end{tabular}
\caption{Memory latency measurements for PicoRV32 as measured using a sequence of 3 instructions. The measured latency is the raw cycle count from the test whereas the true latency is adjusted for cycles spent in the cache hierarchy}
\label{table:mem_latency}
\vspace{-.6cm}
\end{table}

\subsection{Microbenchmarks}
\begin{figure}[t]
\includegraphics[width=3.25in]{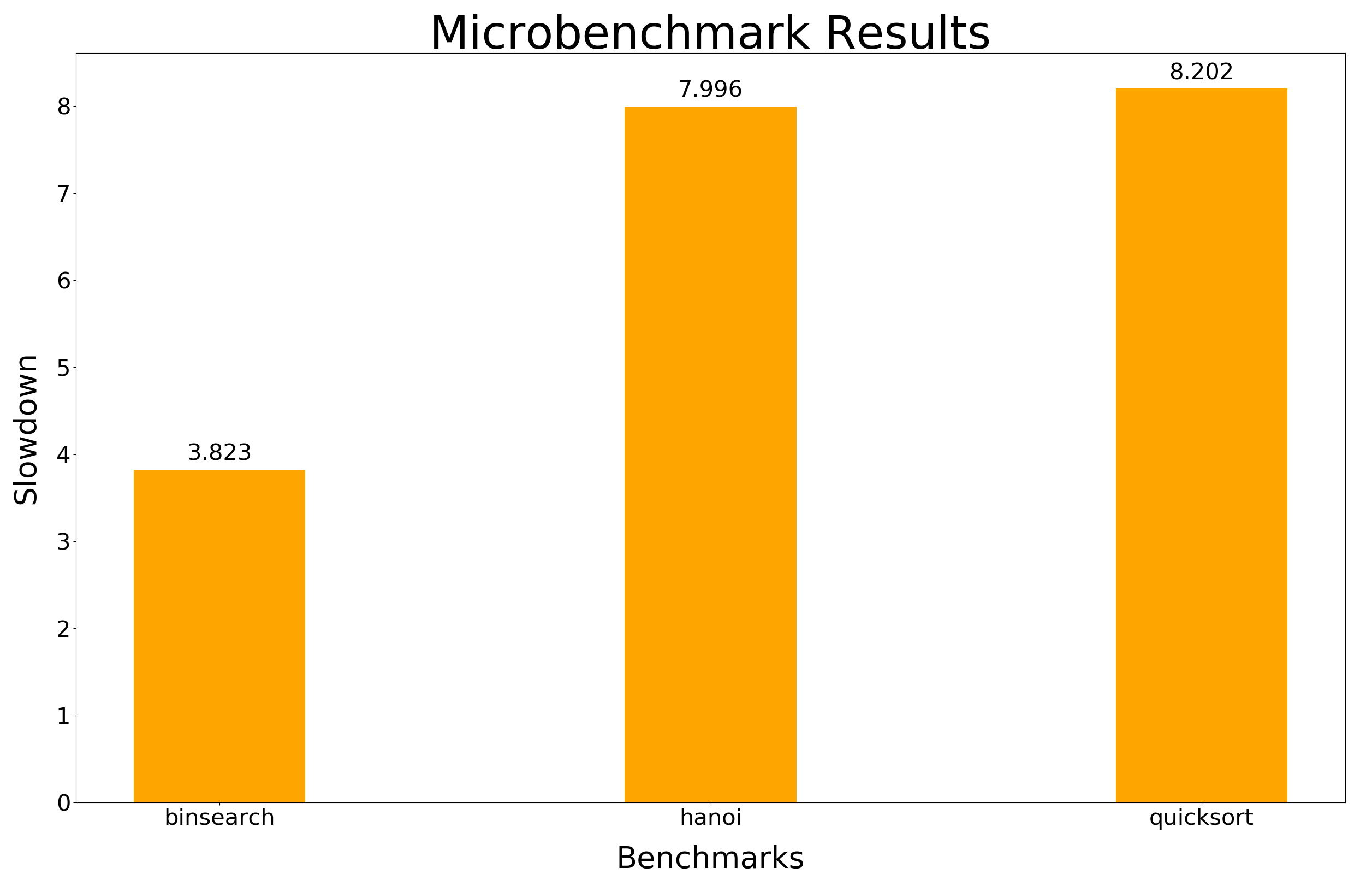}
\caption{Slowdown from running the microbenchmarks on the PicoRV32 core versus the OpenSPARC T1 core. The slowdown values are also given over each bar}
\label{fig:bmark_results}
\vspace{-.5cm}
\end{figure}

We ran three microbenchmarks to compare the performance of the PicoRV32 core integrated into the OpenPiton framework to that of the OpenSPARC T1 core. The bitfile was running at a frequency of 66.667 MHz and had one OpenSPARC T1 tile and one PicoRV32 tile. 
The first microbenchmark was a program that simulated solving the Towers of Hanoi puzzle recursively (\texttt{hanoi}). The second was a binary search program (\texttt{binsearch}), and the third was a quicksort program (\texttt{quicksort}). The Towers of Hanoi is run with a height of 7. The benchmark recursively calls the same function to simulate moving the disks. The binary search benchmark searches for 10 32-bit integer keys randomly chosen in an array of 10,000 32-bit integers. The quicksort benchmark sorts an array of 100 32-bit integers shuffled randomly. 

The slowdown of running each of these benchmarks is shown in Figure \ref{fig:bmark_results}. As expected, all microbenchmarks experienced a slowdown when running on the PicoRV32 core since it is a more simplistic core. \texttt{hanoi} and \texttt{quicksort} both saw about an 8x slowdown. \texttt{binsearch} experienced a smaller slowdown at 4x. 

\texttt{binsearch}'s performance was affected less by running on the PicoRV32 core, because its working set does not fit in the L1.5 cache, which is 8KB. The working set does fit within the L2 cache although there is still the possibility of conflict misses. As a result of the working set size, both cores are forced to access the L2 cache or memory often. Since operations that must go to the L1.5 cache or beyond take approximately the same amount of time for the PicoRV32 core and the OpenSPARC T1 core, \texttt{binsearch} is less impacted by running on the PicoRV32 core.

Although microbenchmarks running on the PicoRV32 suffer reduced performance, the PicoRV32 is designed to minimize area and maximize frequency, essentially trading performance for area and timing. We also expect the PicoRV32 core would consume less energy. 

In our evaluation, the OpenSPARC T1 core and the PicoRV32 core were running at the same clock frequency. To take advantage of the PicoRV32 core's higher maximum frequency, the PicoRV32 core could be put in a different clock domain from the rest of the design and run at a higher frequency to lessen the performance difference. 
	
It is worth noting that the OpenSPARC T1 core was designed for throughput and not single-threaded performance. For example, a thread will be descheduled until a branch is resolved. The core originally had 4 threads to overlap useful work from other threads with long latency instructions.

These trade-offs between performance and other metrics is an intended consequence of having a heterogeneous system architecture. An intelligent scheduler would optimize for these trade-offs and make use of the most appropriate core for its performance and energy-consumption goals.

\section{Related Works}
Kumar et al. explored using multicore processors where cores had heterogeneous microarchitectures but a common ISA \cite{Kumar_power_reduction,Kumar_performance}. Using a variety of simulated cores, they found performance and energy efficiency benefits by scheduling applications on the cores that best match the applications' demands. This work motivated research into heterogeneous architectures, but only looked at cores using one ISA.

Venkat et al. used simulation to explore cores with heterogeneous microarchitecture and heterogeneous ISAs\cite{venkat-hetero-chip}. They used combinations of ARM Thumb, x86, and Alpha cores and found further performance and energy efficiency benefits over just heterogeneous microarchitecture. However, they built their system in simulation, which is of limited usefulness for prototyping hardware and software infrastructure.

The PULP Platform HERO project does provide a heterogeneous-ISA platform \cite{pulp_hero} . They use an ARM core and RISC-V cores. Although the RISC-V cores are implemented on FPGA and can be modified, the ARM core is a hard core. This limits its use for prototyping since the ARM core cannot be modified.

Mantovani et al. implemented an FPGA-based framework for prototyping and analyzing heterogeneous SoCs \cite{carloni_dac,carloni_cases}. However, their focus is on accelerators rather than general-purpose cores.

DeVuyst et al. \cite{DeVuyst-exe-migration} built a compiler and infrastructure for runtime migration in heterogeneous-ISA systems. Using ARM and MIPS cores, they were able to migrate binaries during runtime between cores. Taking advantage of shared memory, they were able to achieve a total performance loss of under 5\% even when migrating every few hundred milliseconds. A key to achieving this performance, however, was the availability of hardware shared memory, so they performed their experiments in simulation. Additionally, they did not use an OS in their evaluation.

The researchers behind Popcorn Linux have also explored building a compiler that allows for runtime migration as well as OS support for heterogeneous-ISA systems \cite{popcorn_mars, popcorn_os_multi_isa}. They used their multikernel model to investigate a potential OS design for a heterogeneous-ISA system  by compiling a copy of their kernel for each ISA. For their evaluations, they used a hardware x86-ARM system where the cores were connected over PCI, but the system did not have hardware shared memory, which meant that migration of binaries during execution was expensive due to the overhead of copying state. JuxtaPiton could provide better insight into the cost of migration of binaries in Popcorn Linux since it has shared memory available.

Lin et al. built K2 OS\cite{k2_os}, an OS which assumes multiple coherence domains where cores in different domains do not have coherent memory. In their hardware model, they assume that cores in different domains can be of different ISAs. Using modified Linux kernels, they run a main kernel in one domain and a shadow kernel in another and replicate state between them. Although K2 is able to run without shared memory, their model supports heterogeneous-ISA cores and could be used in a shared memory system as well. 

\subsection{Enabled research}
Although the PicoRV32 core we incorporated had no caches to simplify interfacing with the L1.5 caches, the same method of adding transducers between a different core's L1 cache and OpenPiton's L1.5 cache could be used to add a more complex core. Our initial experience investigating the integration of more complex cores indicates that this should be relatively straightforward. Other RISC-V cores that could be integrated include "medium" cores such as Ariane \cite{ariane} or Rocket \cite{rocket} or "large" cores such as Anycore \cite{anycore_riscv} or BOOM \cite{boom}.

JuxtaPiton could also be paired with another open-source FPGA framework, like that developed by Mantovani et al. which focuses on accelerators, to create a platform with numerous heterogeneous elements. This would enable researchers to explore heterogeneous architectures with the ability to modify any component of the system and prototype their design on FPGA.

JuxtaPiton can also help enable systems research by providing shared memory on an FPGA. Shared memory is a familiar programming model and allows for efficient migration between cores as found by previous work. At the same time, emulating the design on FPGA enables research into complex, full-stack software that would not be practical in simulation. We expect these unique benefits that JuxtaPiton provides will enable future OS and runtime migration work.

\section{Conclusion}
With an increasing emphasis on energy efficiency in computer systems, it is becoming common to see architectures with heterogeneous processing elements, creating a need for better frameworks for use in research and prototyping. We built JuxtaPiton to enable heterogeneous-ISA research by integrating two open-source projects: OpenPiton and PicoRV32. JuxtaPiton is the first time a new core has been integrated into the OpenPiton framework, and we belive it is the first open-source, general-purpose, heterogeneous-ISA processor.
We evaluated trade-offs of using the PicoRV32 core or the OpenSPARC T1 core. We found that although the PicoRV32 core experienced a slowdown in the microbenchmarks we ran, it used much less area than the OpenSPARC T1 core and could improve timing in the design. 
We believe that this FPGA implementation of a heterogeneous-ISA, shared memory, multiprocessor will enable future research. Architects will be able to modify it and prototype their designs on FPGA. OS researchers will be able to evaluate more complex software designs on realistic hardware prototypes while also taking advantage of shared memory.

\begin{acks}
This work was partially supported by the NSF under Grants No. CNS-1823222, CCF-1217553, CCF- 1453112, and CCF-1438980, AFOSR under Grant No. FA9550-14-1-0148, AFRL and DARPA under agreements No. FA8650-18-2-7846 and FA8650-18-2-7852, and DARPA under Grants No. N66001-14-1-4040 and HR0011-13-2-0005. The U.S. Government is authorized to reproduce and distribute reprints for Governmental purposes notwithstanding any copyright notation thereon. The views and conclusions contained herein are those of the authors and should not be interpreted as necessarily representing the official policies or endorsements, either expressed or implied, of Air Force Research Laboratory (AFRL) and Defense Advanced Research Projects Agency (DARPA), the NSF, AFOSR, DARPA, or the U.S. Government.
\end{acks}

\bibliographystyle{ACM-Reference-Format}
\bibliography{references}

\end{document}